\documentclass[twocolumn,aps,prl,superscriptaddress]{revtex4-1}
\usepackage{graphicx,subfigure}
\usepackage{amsmath,amssymb,bm}
\usepackage{mathtools}
\usepackage{multirow}
\usepackage{makecell}
\usepackage{textcomp}
\usepackage{float}
\usepackage{xcolor, soul}

\usepackage{color}
\usepackage[normalem]{ulem}
\usepackage{dsfont}
\usepackage{mathrsfs}
\usepackage{braket}
\usepackage{transparent}
\usepackage{xcolor}
\usepackage{hhline}
\usepackage{graphicx}
\usepackage{pdfpages}
\usepackage{bibunits}
\usepackage{filecontents,hyperref}

\makeatletter
\AtBeginDocument{\let\LS@rot\@undefined}
\makeatother

\usepackage{graphicx}

\graphicspath{ {figures/} }

\makeatletter
\let\saved@includegraphics\includegraphics
\AtBeginDocument{\let\includegraphics\saved@includegraphics}
\makeatother

\newcommand{\figOne}{
 \begin{figure}[t]
    \centering
    \includegraphics[width=\columnwidth]{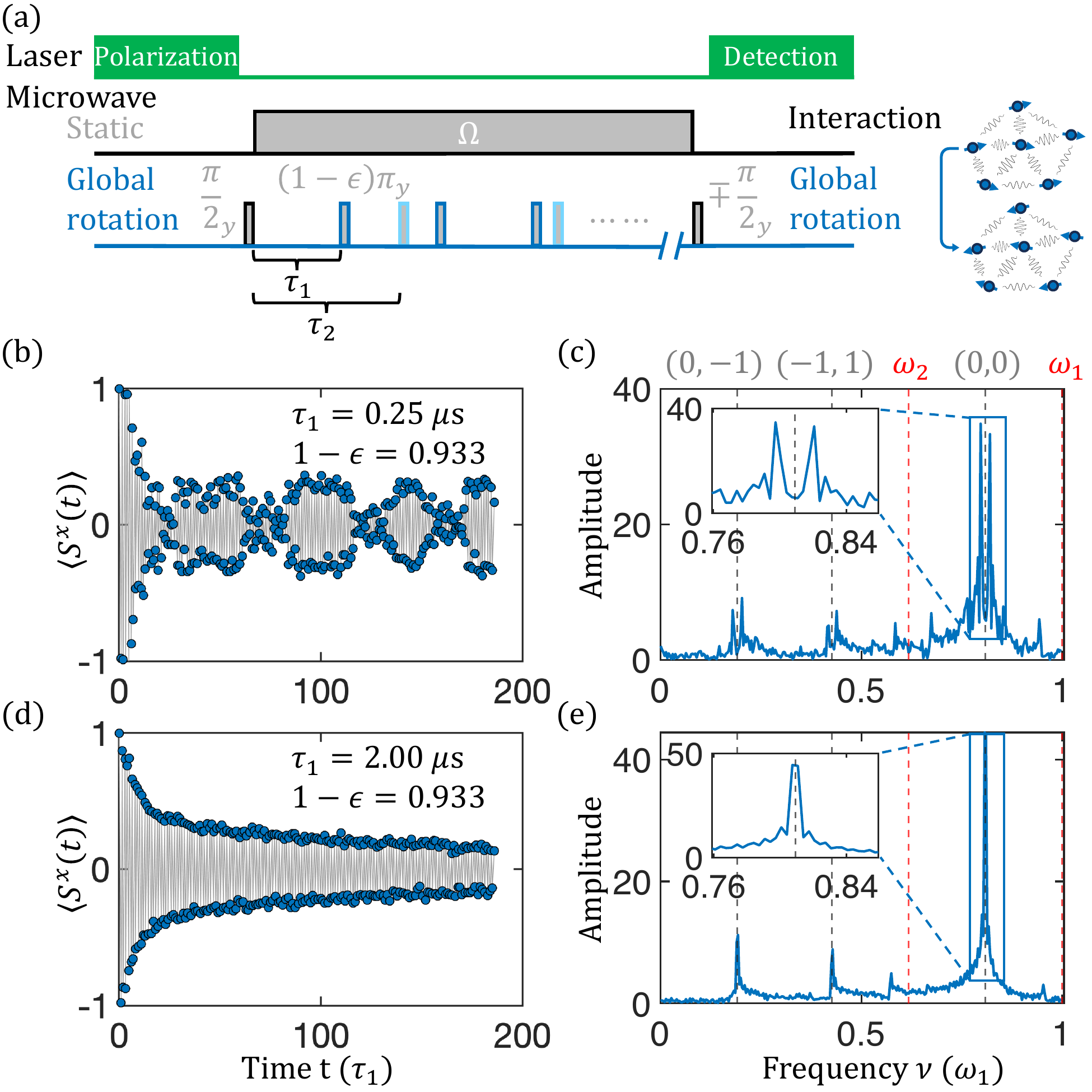}
    \caption{{\bf Experiment to observe $\mathbb{Z}_2$ time quasi-crystalline order.} (a) Pulse sequence for the experiment. After laser polarization, a $\frac{\pi}{2}$-pulse along $\hat{y}$-axis initializes the spin ensemble to $|+x\rangle$, followed by a time-independent static microwave drive $\Omega \sum S^x_i$. The quasi-periodic drives consist of a series of global pulses with rotation angle $(1-\epsilon)\pi$ applied with incommensurate frequencies $\omega_1 = 2\pi/\tau_1$ and $\omega_2 = 2\pi/\tau_2$. A final $\mp\frac{\pi}{2}$-pulse along $\hat{y}$-axis flips the spins back to $\hat{z}$ basis for measurement. (b)-(e) Representative spin polarization dynamics and the corresponding Fourier spectrum under quasi-periodic drive with [$\tau_1 = 0.25~\mu$s, $1-\epsilon = 0.933$, (b)-(c)] and [$\tau_1 =2.00~\mu$s, $1-\epsilon = 0.933$, (d)-(e)]. Gray dashed lines, labeled by $(N_1,N_2)$, indicate the sub-harmonic response frequencies of DTQC. Red dashed lines mark the quasi-periodic driving frequencies. Insets: zoom in on the sub-harmonic response at $\frac{1}{2}(\omega_1+\omega_2)$.
    }
    \label{fig:fig1}
\end{figure}
}

\newcommand{\figTwo}{
  \begin{figure*}[t]
    \centering
    \includegraphics[width=0.7\textwidth]{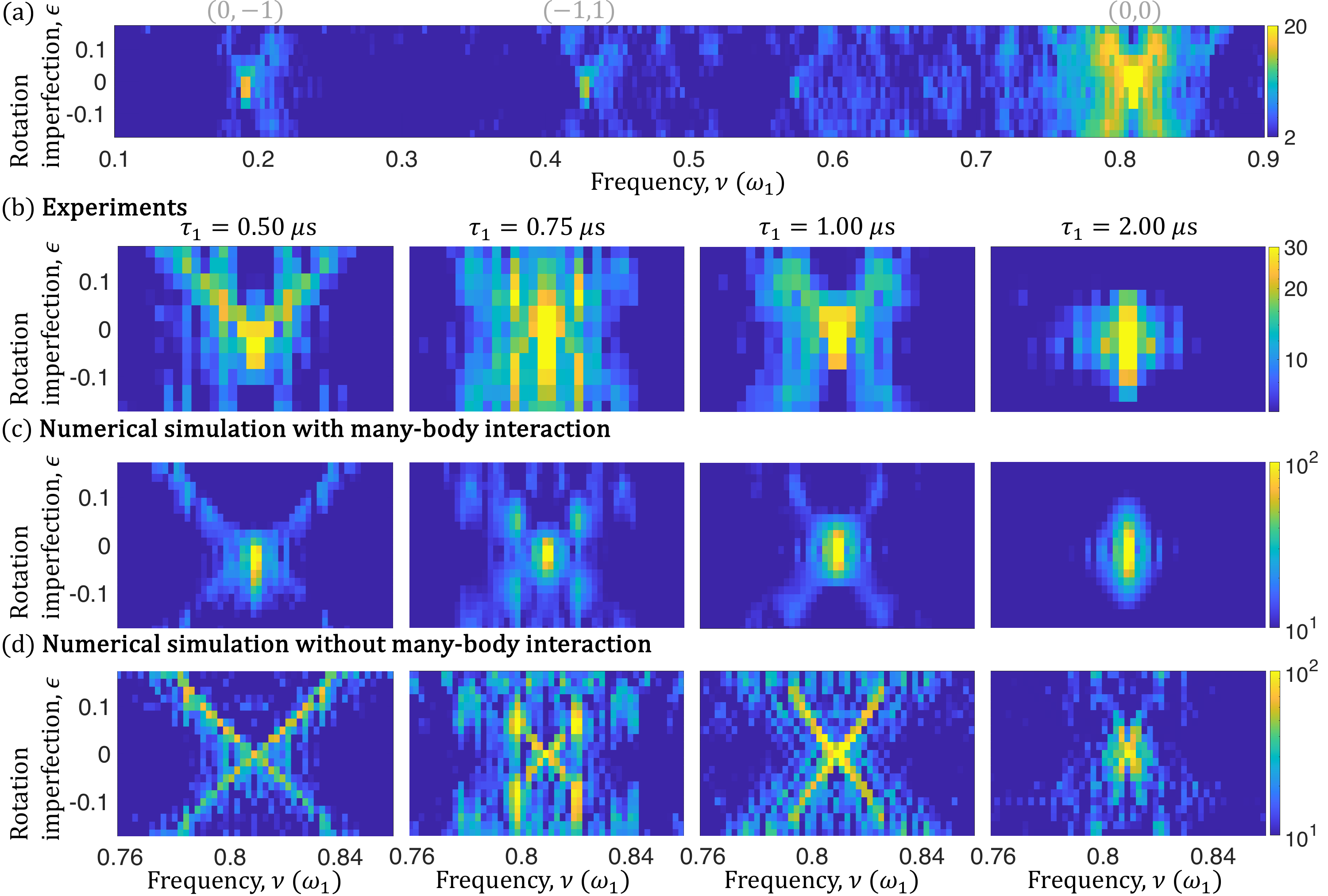}
    \caption{{\bf Rigidity of time quasi-crystals.} (a) Fourier spectrum of DTQC at $\tau_1 = 1.00~\mu$s with varying rotation imperfections. All three sub-harmonic responses remain robust when $\epsilon \in [-0.07, 0.05]$, and split beyond this regime. Color bar indicates amplitude of the Fourier transform. (b)-(d) A comparison between experimental data and numerical simulation [zooming in on the response at $\frac{1}{2}(\omega_1+\omega_2)$]. We observe quantitative agreement between experiments [(b)] and many-body simulations using 16 disordered spins [(c)]. The numerical simulation without many-body interaction [(d)] indicates that any small rotation imperfection induces splitting of the sub-harmonic responses which does not agree with our experimental observations.
    }
    \label{fig:fig2}
\end{figure*}
}

\newcommand{\figThree}{
\begin{figure}[t]
    \centering
    \includegraphics[width=\columnwidth]{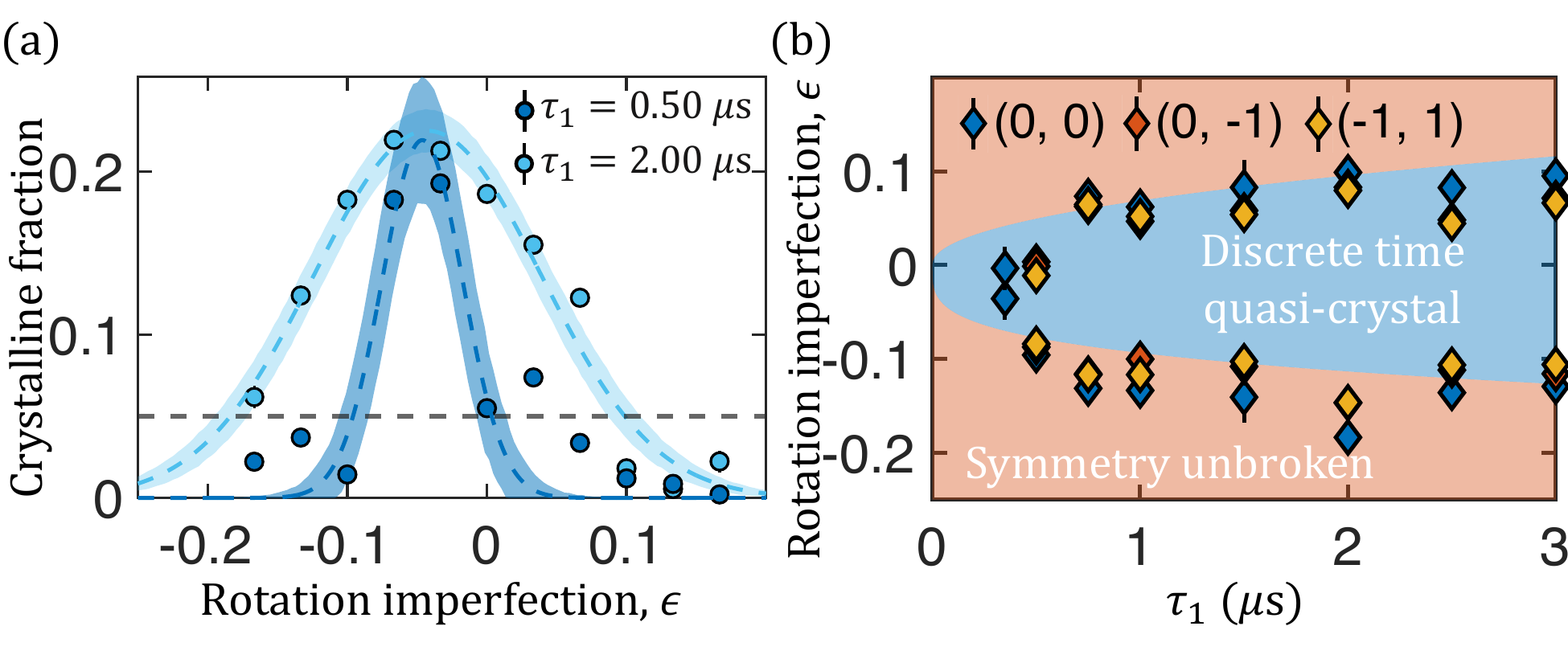}
    \caption{{\bf Phase diagram.} (a) Crystalline fraction, $f(\nu)$, as a function of rotation imperfection for (0,0) sub-harmonic response. Dashed blue curves are Gaussian fits to the data, and shaded area indicates fitting uncertainties accounting for 1 standard deviation. Phase boundary are phenomenologically defined as $5\%$ of the crystalline fraction, marked by the horizontal gray dashed line. The phase boundary extends at larger interaction time (stronger interaction).
    (b) The phase diagram extracted using three distinct sub-harmonic responses (0, 0) [blue], (0, -1) [red], and (-1, 1) [yellow] are consistent with each other. The boundaries extend with increasing $\tau_1$, and saturate around $2~\mu$s. Background colors mark the phase boundary obtained from the many-body numerical simulation using N=16 disordered spins.
    }
    \label{fig:fig3}
\end{figure}
}

\newcommand{\figFour}{
\begin{figure*}[t]
    \centering
    \includegraphics[width=0.7\textwidth]{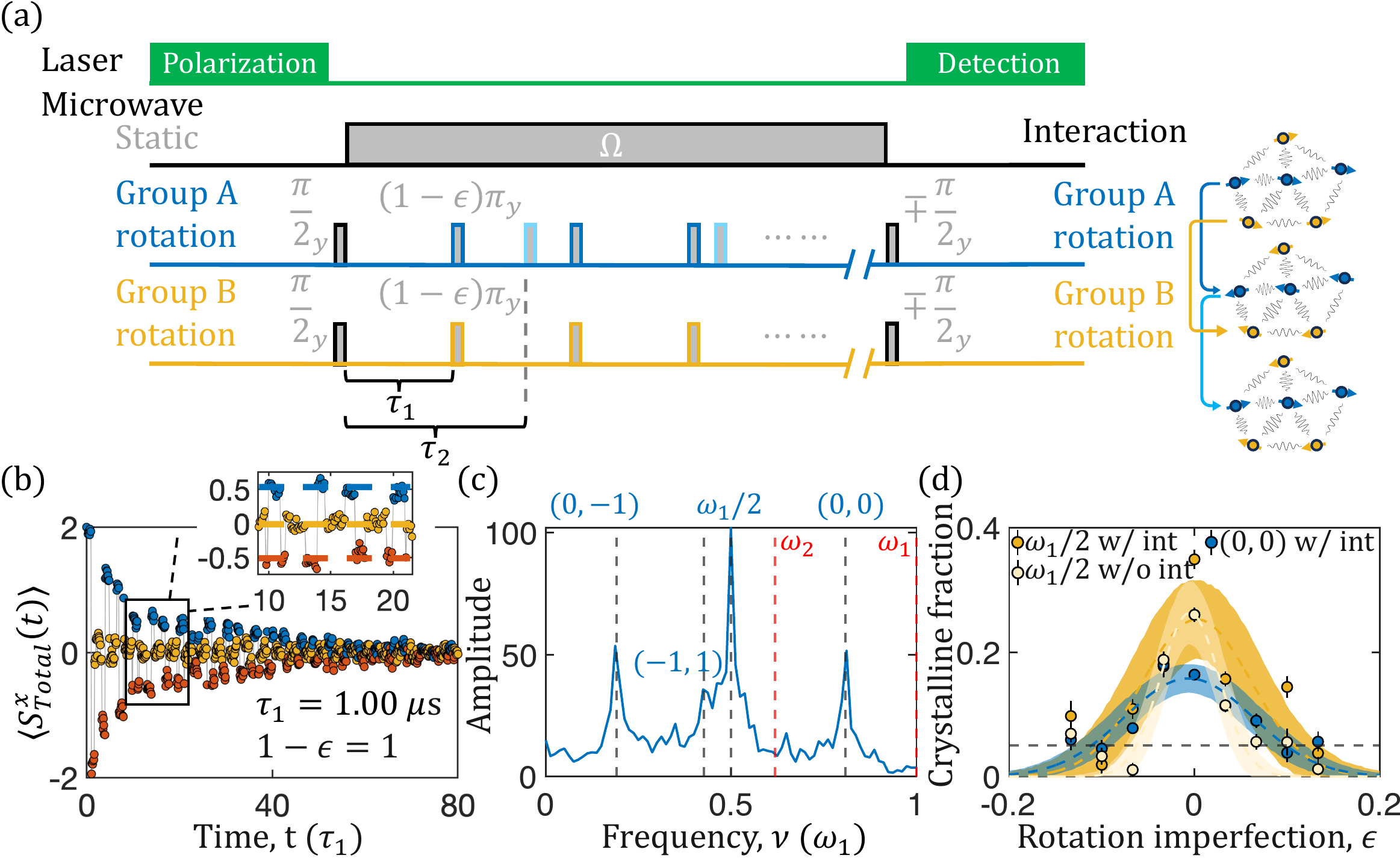}
    \caption{{\bf Observing $\mathbb{Z}_2\times \mathbb{Z}_2$ time quasi-crystalline order.} (a) Experimental pulse sequence. Global rotations are applied at times $t = n_1\tau_1$ for both NV groups, and at times $t = n_2\tau_2$ only for group B. The total spin polarization for both NV groups is measured at the end. (b) Representative polarization dynamics with $\tau_1 = 1~\mu$s and $\epsilon = 0$. The sampling rate is set to be $10\omega_1$. Markers change colors as a rotation pulse is applied. Inset: a zoom in of the polarization dynamics oscillating among three discrete values.
    (c) Fourier spectrum of the total polarization dynamics. Four dominant sub-harmonic responses at frequencies $\{\frac{1}{2}(\omega_1+\omega_2),~ \frac{1}{2}(\omega_1-\omega_2),~\frac{1}{2}(-\omega_1+3\omega_2),~\frac{1}{2}\omega_1 \}$ are marked with blue dashed lines. Red dashed lines are the quasi-periodic driving frequencies. (d) The extracted crystalline fraction at $\frac{1}{2}\omega_1$ with (dark yellow) and without (light yellow) inter-group interaction for $\tau_1 = 1.00~\mu$s. We also examine the crystalline fraction at $\frac{1}{2}(\omega_1+\omega_2)$ with inter-group interaction (blue), whose width is consistent with the $\frac{1}{2}\omega_1$ response. Dashed lines are Gaussian fits and shaded area indicates fitting uncertainties accounting for 1 standard deviation.
    }
    \label{fig:fig4}
\end{figure*}
}

\makeatletter
\newcommand*{\centerfloat}{%
  \parindent \z@
  \leftskip \z@ \@plus 1fil \@minus \textwidth
  \rightskip\leftskip
  \parfillskip \z@skip}
\makeatother

\usepackage{lineno} 
\begin{document}

\title{Experimental Realization of Discrete Time Quasi-Crystals}

\author{Guanghui~He,$^{1,*}$
Bingtian~Ye,$^{2,*,\dag}$
Ruotian~Gong,$^{1}$
Changyu~Yao,$^{1}$
Zhongyuan~Liu,$^{1}$\\
Kater~W.~Murch,$^{1,3}$
Norman~Y.~Yao,$^{2}$
Chong~Zu,$^{1,3,4,\dag}$
\\
\normalsize{$^{1}$Department of Physics, Washington University, St. Louis, MO 63130, USA}\\
\normalsize{$^{2}$Department of Physics, Harvard University, Cambridge, MA 02138, USA}\\
\normalsize{$^{3}$Center for Quantum Leaps, Washington University, St. Louis, MO 63130, USA}\\
\normalsize{$^{4}$Institute of Materials Science and Engineering, Washington University, St. Louis, MO 63130, USA}\\
\normalsize{$^*$These authors contributed equally to this work.}\\
\normalsize{$^\dag$To whom correspondence should be addressed: bingtian\_ye@g.harvard.edu, zu@wustl.edu}\\
}

\begin{abstract}
Floquet (periodically driven) systems can give rise to unique non-equilibrium phases of matter without equilibrium analogs.
The most prominent example is the realization of discrete time crystals.
An intriguing question emerges: what other novel phases can manifest when the constraint of time periodicity is relaxed? In this study, we explore quantum systems subjected to a quasi-periodic drive. 
Leveraging a strongly interacting spin ensemble in diamond, we identify the emergence of long-lived discrete time quasi-crystals. 
Unlike conventional time crystals, time quasi-crystals exhibit robust sub-harmonic responses at multiple incommensurate frequencies. 
Furthermore, we show that the multi-frequency nature of the quasi-periodic drive allows for the formation of diverse patterns associated with different discrete time quasi-crystalline phases. 
Our findings demonstrate the existence of non-equilibrium phases in quasi-Floquet settings, significantly broadening the catalog of novel phenomena in driven many-body quantum systems.

\end{abstract}
\date{\today}
\maketitle

\emph{Introduction} --- Spontaneous symmetry breaking is fundamental in all fields of physical sciences \cite{chaikin1995principles}.
In the context of many-body physics, the emergence of distinct phases can be viewed through the lens of different types of symmetry breaking \cite{sachdev1999quantum}.
The most prominent example is probably the formation of crystals from breaking the translational symmetry in space.
Its cousin in the time domain, the recently discovered discrete time crystal (DTC), breaks the time-translation symmetry \cite{wilczek2012quantum,wilczek2013superfluidity,else2020discrete,khemani2016phase,von2016phase,else2016floquet,else2017prethermal,ho2017critical,moessner2017equilibration,harper2020topology,yao2017discrete,yao2020classical,mi2022time,lazarides2017fate,frey2022realization,rovny2018observation,yang2021dynamical,kessler2020continuous,taheri2022all,kuros2020phase,yu2019discrete,gambetta2019classical,ippoliti2021many,gambetta2019discrete,liu2023discrete,ye2021floquet,kongkhambut2022observation,autti2018observation,kessler2021observation,beatrez2023critical, zaletel2023colloquium, xiang2024long, huang2018clean}.
For quite a long time, periodicity and long-range order are believed to be the two intimately correlated characteristics of crystalline phases \cite{choi2017observation,zhang2017observation,zaletel2023colloquium}.
Quasi-crystals are an exception: they are ordered but not apparently periodic \cite{else2020long}.
The hidden periodicity can be thought of as a higher-dimensional periodic crystal projecting onto a lower-dimensional subspace. 
A natural question arises as whether one can make an analogy in the time-domain to create time quasi-crystalline orders.

To answer this question, one can first draw intuition from the creation of a conventional time crystal.
Since the quantum equilibrium state of a time-independent Hamiltonian cannot have time-dependent observables, the construction of a time crystal is only possible in systems with discrete time-translation symmetry, i.e. in a periodically driven (Floquet) system \cite{bruno2012comment,bruno2013impossibility,watanabe2015absence}.

The construction of DTC relies on the concept of emergent symmetry which does not exist in the native time-dependent Hamiltonian; instead it is induced by the time-translation symmetry. 
A characteristic of the DTC phase is that the physical observable exhibits a longer period than the driving field \cite{sacha2015modeling}.
A prototypical example is the dynamical decoupling sequence with a train of evenly-spaced $\pi$-pulses, where the effective Hamiltonian in the ``toggling" frame exhibits a $\mathbb{Z}_2$ Ising symmetry that is independent of the symmetry of the initial Hamiltonian.
The breaking of such emergent symmetry then enables the system to display a sub-harmonic response.
Crucially, as a phase of matter (in contrast to fine-tuned Floquet engineering techniques), the key ingredient is the robustness of the above-mentioned sub-harmonic response to perturbations.
To go beyond DTC and create a discrete time quasi-crystal (DTQC), a straightforward idea is to replace the periodically applied $\pi$-pulses with a quasi-periodic manner, where the rigidity of the crystalline order becomes a more complex question.

In this Letter, we investigate the DTQC order using a strongly interacting ensemble of nitrogen-vacancy (NV) centers in diamond.
Our results are in three-fold.
First, we observe the DTQC order manifests as the responses at halves of the linear combinations of two incommensurate driving frequencies (Fig.~\ref{fig:fig1}). 
Second, with regard to the robustness of the DTQC order, we find that the strong many-body interaction within the spin ensemble provides rigidity against perturbations (Fig.~\ref{fig:fig2}). 
We quantify this rigidity by varying the competition between the interaction strength and the magnitude of the perturbation (controlled by the applied pulse durations) to map out the phase diagram of DTQC (Fig.~\ref{fig:fig3}).
Third, we show that the multifrequency nature of the quasi-periodic drives allows one to incorporate multiple symmetries into the effective Hamiltonian and realize more complicated DTQCs.
For instance, by leveraging two different crystallographic groups of NV centers, we investigate the $\mathbb{Z}_2\times\mathbb{Z}_2$ DTQC phase (Fig.~\ref{fig:fig4}).

\emph{Observing DTQC order} --- To experimentally realize DTQC, we refine the aforementioned intuition into a concrete quasi-Floquet Hamiltonian,
\begin{equation} \label{eq1}
\begin{split}
\mathcal{H}(t) &= \sum_{i,j} (J_{i,j} S^x_i S^x_j) +  \Omega \sum_{i} S^x_i \\
&+ (1-\epsilon)\sum_{i} S_i^y \left[\sum_{n_1}\delta(t-n_1\tau_\mathrm{1}) + \sum_{n_2}\delta(t-n_2\tau_\mathrm{2})\right],
\end{split}
\end{equation}
where $\hat{S}$ is the spin operator, $n_1, n_2 \in \mathbb{Z}$, and $|\epsilon|\ll1$ represents a small perturbation on top of perfect $\pi$-pulse rotations.
The first two terms correspond to the time-independent component of the many-body spin Hamiltonian, and the last term is the quasi-periodic spin flip performed with frequencies $\omega_1 = 2\pi/\tau_1$ and $\omega_2 = 2\pi/\tau_2$.
We choose the ratio $\varphi = \omega_1/\omega_2 = (\sqrt{5}+1)/2\approx 1.618$ as the golden ratio so that the two driving frequencies are the most incommensurate to each other.

To implement the quasi-Floquet Hamiltonian, we choose to work with  a diamond sample containing a dense ensemble of NV centers ($\sim 4.5~$ppm). 
Each NV center is a spin-1 artificial atom that can be optically initialized and readout.
By applying an external magnetic field, we lift the degeneracy of $|m_s = \pm1\rangle$ via Zeeman splitting, and isolate an effective spin-$1/2$ system spanned by $\{|m_s = 0\rangle, |m_s = -1\rangle\}$. 
Under a static driving field along the $\hat{x}$ direction, the Hamiltonian governing the dipolar interacting NV ensemble takes the form \begin{equation} \label{eq2}
\begin{split}
 \mathcal{H}_{0} = &-\sum_{i,j} J_{i,j}(S^z_i S^z_j - S^x_i S^x_j - S^y_i S^y_j) \\
&+\sum_{i} h_i S_i^z + \Omega\sum_{i} S^x_i,
\end{split}
\end{equation}
where $J_{i,j} \sim (2\pi)\times52~$MHz$\cdot$nm$^3/r_{i,j}^3$ with $r_{i,j}$ the distance between the $i^{th}$ and $j^{th}$ spins, $h_i$ characterizes the on-site random field, $\Omega$ is the Rabi frequency of the drive.

A few remarks are in order.
First, we set the microwave strength [$\Omega= (2\pi)\times 8.3~$MHz] to be much larger than both the average dipolar interaction strength between two nearby NVs [$\overline{J}_{i,j} \approx (2\pi)\times 0.05~$MHz] and the magnitude of the on-site random field [$\overline{h}_{i} \approx (2\pi)\times 1.7$~MHz], so that all the terms that do not commute with $S_i^x$ are significantly suppressed.
In this limit, the many-body Hamiltonian in Eqn.~\ref{eq2} can be well-approximated by $\mathcal{H}_{0} \approx \sum_{i<j} (J_{i,j} S^x_i S^x_j) +  \Omega \sum_{i} S^x_i$, corresponding to the first two terms of our quasi-Floquet Hamiltonian (Eqn.~\ref{eq1}).
Second, the quasi-periodic driving terms of the Hamiltonian can be realized via series of global microwave pulses along $\hat{y}$ axis with rotation angle $(1-\epsilon)\pi$ at times $t = n_1\tau_1$ and $n_2 \tau_2$.
Third, the random positioning of the NV centers can introduce strong disorder to the spin-spin interaction term, $J_{i,j}$, leading to slow critical thermalization of the spin dynamics as demonstrated in previous works \cite{kucsko2018critical, choi2017observation}.
Therefore, although the DTQC order studied in this work does not exhibit infinitely long lifetime due to the absence of many-body localization,
strong disorder still prevents the system from approaching a prethermal state~\footnote{The angular dependence of the dipolar interaction, $J_{i,j}$, can be both positive and negative. For a polarized initial state, the effective temperature is close to infinity, so that the prethermal state is featureless and cannot support order \cite{he2023quasi,vu2023dissipative}} and enables a long-lived DTQC order within the relevant times investigated in experiment.

\figOne

\figTwo

With all the ingredients in hands, we now turn to the full experimental sequence shown in Fig.~\ref{fig:fig1}a. 
After the optical pumping, the NV system is initialized to $|+x\rangle = \otimes_{i} \frac{|0\rangle_i+|-1\rangle_i}{\sqrt{2}}$ by applying a global $\frac{\pi}{2}$ pulse around the $\hat{y}$ axis, followed by the quasi-periodic drives.
Let us first consider a simple scenario.
In the case of $\epsilon \approx 0$, each microwave pulse precisely rotates the spin ensemble half way around the Bloch sphere at integer times of $\tau_1$ and $\tau_2$,
so the sign of the average magnetization along $\hat{x}$ axis, $\langle S^x(t)\rangle$, will flip at the corresponding times, while the amplitude will remain constant in principle, but subject to a slow decay due to the combination of pulse imperfections as well as intrinsic spin relaxation ($T_1$) process in the experiment.
In the Fourier space, the measured magnetization dynamics manifests a series of ``subharmonic" peaks at frequencies~\footnote{Although neither the Hamiltonian nor the dynamics exhibits an exact periodicity, the power spectrum in the Fourier space can be thought of as a projection from the higher-dimensional Fourier spectrum. From this perspective, the dynamics of an observable is effectively a subharmonic response to the higher dimensional Fourier spectrum of the drive.}:
\begin{equation} \label{eq3}
\begin{split}
\nu = (N_1+\frac{1}{2})\omega_1+(N_2+\frac{1}{2})\omega_2, ~~ N_1, N_2\in \mathbb{Z}.
\end{split}
\end{equation}

One may naively consider such dynamics as the DTQC order.
However, the aforementioned picture relies on the fine-tuning of the experimental parameters, i.e.~ the rotation pulses need to be perfect.
In order to establish a real phase of matter, we have to investigate the rigidity of the ``subharmonic" responses against the imperfections of the $\pi-$rotations. 

For relatively short interaction time~\footnote{The effective interaction strength in quasi-Floquet cycles can be tuned by varying the interaction time $\tau$ and $\varphi\tau$~\cite{choi2017observation}}, $\tau_1 = 0.25~\mu$s and $\tau_2 = 0.404~\mu$s, in the presence of pulse imperfections, $1-\epsilon\approx 0.933$, we observe a clear beating in the measured spin dynamics (Fig.~\ref{fig:fig1}b).
In the corresponding Fourier spectrum, we observe that the three dominant sub-harmonic responses marked with ($N_1, N_2$) at (0, 0), (0, -1), (-1, 1), each splits into two peaks (Fig.~\ref{fig:fig1}c), suggesting the breakdown of the DTQC order.
We note that, in the presence of quasi-periodic rotations, there does not exist a stroboscopic time that one can measure the spin dynamics. 
As a result, we can probe the spin dynamics with a much faster sampling frequency, $\omega_\mathrm{s}=10\omega_1$, or shorter time-interval, $\tau_1/10$.
However, we also verify that when sampling at the quasi-periodic time, $\tau = n_1 \tau_1$ and $n_2 \tau_2$, one can still faithfully obtain the spin dynamics and the associated Fourier spectrum (see Supplementary Materials).

Surprisingly, for long interaction time ($\tau_1 = 2~\mu$s, and $\tau_2 = 3.236~\mu$s) with the same magnitude of the pulse imperfections, the measured dynamics remains ordered up to time $t\sim 200\tau_1$ (Fig.~\ref{fig:fig1}d).
The robustness is also evident in the frequency domain --- the three dominant ``subharmonic" peaks do not split.
The distinct dynamical behaviors with different interaction times suggests that the many-body interaction plays a crucial role in stabilizing the DTQC phase.
Intuitively, the coupling terms in the Hamiltonian, $J_{i,j}S_i^x S_j^x$, provide an energy barrier that prevents the uncorrelated flips of each individual spin caused by the local perturbations, i.e. the imperfect pulse rotation in experiment.

Such intuition immediately suggests that, given an interaction time $\tau_1$, the DTQC behavior should be rigid against imperfections up to certain magnitudes.
This is indeed borne out by our data. At $\tau_1 = 1~\mu$s, when $\epsilon \in [-0.07, 0.05]$, all the three ``sub-harmonic" responses remain rigid (Fig.~\ref{fig:fig2}a).
When the imperfection goes beyond this regime, all the three peaks split, indicating the breakdown of DTQC order. 

\emph{Rigidity of DTQC order} --- To further corroborate that the rigidity of DTQC order originates from the many-body interaction between spins, $J_{i,j}S_i^x S_j^x$, we compare our experimental data to the numerical simulations with and without the interaction using 16 randomly positioned NV spins (see Supplementary Materials).
Let us focus on the most prominent ``sub-harmonic" peak, (0,0). 
In the absence of many-body interactions, any small $\epsilon$ will lead to the splitting of the peaks across all interaction times $\tau_1$ (Fig.~\ref{fig:fig2}d), in sharp contrast to the experimental observations (Fig.~\ref{fig:fig2}b).
When adding the interactions into the system Hamiltonian, the numerics can reproduce the robustness of DTQC order observed in experiment (Fig.~\ref{fig:fig2}c).

Crucially, when the interaction strength increases, we observe that the DTQC order is more resilient to the magnitude of the imperfections, $\epsilon$.
To quantify this, we first define a crystalline fraction,
\begin{equation} \label{eq4}
\begin{split}
f(\epsilon) = \frac{|S(\nu_0)|}{\displaystyle\sum_{\nu_0-\delta}^{\nu_0+\delta}|S(\nu)|},
\end{split}
\end{equation}
where $\nu_0$ denotes the frequency of the sub-harmonic peak labeled by ($N_1=0$, $N_2=0$), and $\delta = 0.1\omega_1$ so that it does not overlap with the other peaks.
As expected, with strong interaction ($\tau_1 = 2~\mu$s) the crystalline fraction vanishes at a significantly larger imperfection value compared to the small interaction case ($\tau_1 = 0.5~\mu$s) (Fig.~\ref{fig:fig3}a).

To map out the phase boundary for the DTQC order, we determine the critical imperfection $\epsilon_c(\tau_1)$ such that the crystalline fraction meets the threshold $f(\epsilon_c)=0.05$. 
Fig.~\ref{fig:fig3}b shows the extracted phase diagram of the DTQC order. 
We apply the same strategy to the other two sub-harmonic harmonics responses, as well as the numerical simulations.
Interestingly, the phase boundaries determined by the distinct ``sub-harmonic" responses are consistent with each other, and also agree well with the numerics.
The co-existence of the ``sub-harmonic" peaks marks an unique feature of the DTQC phase.

\figThree

A few points need to be addressed.
First, in both experiment and simulation, the phase boundary saturates around $\tau_1 \gtrsim 2~\mu$s, suggesting that the rigidity of DTQC can not be further enhanced by simply extending the interaction time $\tau_1$.
Stabilizing DTQC order until arbitrarily late time requires many-body localization \cite{basko2006metal,nandkishore2015many,abanin2015exponentially,schreiber2015observation,smith2016many,randall2021many}.
However MBL is believed to not exist in 3D disordered dipolar spin ensembles.
An alternative solution is Floquet prethermalization \cite{peng2021floquet, he2023quasi}. 
While such phenomenon in general can be applied to our system, it requires the evolution to start from a low-temperature initial state, which is not satisfied in 3D dipolar spin ensemble.
As a result, the DTQC order demonstrated in this work still suffers from thermalization; nevertheless, it is suppressed by the presence of strong disorder, giving rise to the long lifetime of the observed DTQC behavior ~\cite{kucsko2018critical}.

Second, at certain interaction time and imperfection range (e.g. at $\tau_1 = 0.75~\mu$s and $\epsilon\in[0.05, 0.12]$), we observe that the splitting of the ``sub-harmonic" response exhibits a few segmented plateaus (Fig.~\ref{fig:fig2}b-d).
Such feature is present even in the non-interacting scenarios, and we attribute it to the complex interplay between the strong locking field $\Omega$ and the quasi-periodic rotation pulses.
Third, one may naively think that the crystalline fraction, $f(\epsilon)$, should be symmetric around $\epsilon = 0$.
However, in both experiment and simulations, we find that the maximum value of $f(\epsilon)$ is slightly shifted to $\epsilon\approx-0.04$ (Fig.~\ref{fig:fig3}).
This is due to the presence of on-site random field, $\sum_i h_i S_i^z$, which leads to a faster spin rotation with an effective Rabi frequency, $\Omega^\mathrm{eff} = \sqrt{\Omega^2+\bar{h}_i^2}$. 

\figFour

\emph{$\mathbb{Z}_2\times \mathbb{Z}_2$ DTQC} --- 
The aforementioned DTQC order is featured by the measured total polarization oscillating between two distinct values at ``sub-harmonic" frequencies, reflecting the underlying emergent $\mathbb{Z}_2$ Ising symmetry along $\hat{x}$.
Here, utilizing two crystallographic groups of NV centers, we turn to constructing a more complex DTQC with $\mathbb{Z}_2\times \mathbb{Z}_2$ symmetry.
In particular, the interaction Hamiltonian including two groups of NV centers modifies Eqn.~\ref{eq1} to:
\begin{equation} \label{eq5}
\begin{split}
\mathcal{H}(t) &= \sum_{\mathclap{\substack{i,j \in A \\ \mathrm{or}~i,j \in B}}} (J_{i,j} S^{x}_{i} S^{x}_{j}) +  \sum_{\mathclap{\substack{i \in A \\ \mathrm{and}~j \in B}}}  (J'_{i,j}S^z_{i} S^z_{j}) +  \Omega \sum_{i\in A~\mathrm{and}~B} S^x_{i} \\
&+ (1-\epsilon)\sum_{i\in A~\mathrm{and}~B} S_i^y \left[\sum_{n_1}\delta(t-n_1\tau_\mathrm{1})\right] \\
&+ (1-\epsilon)\sum_{i\in A} S_i^y \left[\sum_{n_1}\delta(t-n_2\tau_\mathrm{2})\right].
\end{split}
\end{equation}
Here $A$ and $B$ refer to two different groups of NV centers, between which the inter-group interaction is governed by the Ising term.
We choose the static Rabi driving strength $\Omega$ to be the same for group $A$ and $B$, so that the inter-group interaction becomes resonant (see Supplementary Materials).
We perform global rotations along $\hat{y}$ axis for both groups at time $n_1 \tau_1$, and a rotation only for group A at time $n_2 \tau_2$.
The underlying emergent symmetry of the system is expanded to $\mathbb{Z}_2\times \mathbb{Z}_2$, where the two $\mathbb{Z}_2$ symmetries correspond to spin rotations for both groups and a single group A respectively (Fig.~\ref{fig:fig4}a).

Next, we investigate the subharmonic response of the resulting $\mathbb{Z}_2\times \mathbb{Z}_2$ DTQC by probing the total polarization dynamics of both groups.
A representative spin dynamics with $\tau_1=1~\mu$s and $\epsilon=0$ is shown in Fig.~\ref{fig:fig4}b.
In sharp contrast to the $\mathbb{Z}_2$ scenario, we observe the spin dynamics oscillates among three discrete values. 
Since the rotations for two NV groups are performed at  incommensurate times, there is an additional zero polarization when two groups of NV centers are anti-aligned. 
Although the order parameter appears to be zero, the system still remains ordered during these specific time segments, rather than reaching to an equilibrium state. 
The corresponding Fourier spectrum also reflects the same physics (Fig.~\ref{fig:fig4}c): There exist four dominant sub-harmonic peaks at $\{\frac{1}{2}(\omega_1+\omega_2),~ \frac{1}{2}(\omega_1-\omega_2),~\frac{1}{2}(-\omega_1+3\omega_2),~\frac{1}{2}\omega_1 \}$.

To explore the effect of inter-group interaction on the stability of $\mathbb{Z}_2\times \mathbb{Z}_2$ DTQC, we analyze the corresponding DTQC phase boundary with rotation imperfections. 
As shown in Fig.~\ref{fig:fig4}d, with interactions, the crystalline fraction at sub-harmonic frequency $\frac{1}{2}\omega_1$ vanishes at relatively larger imperfection values compared to the non-interacting case, suggesting that the DTQC phase could be more robust in the presence of inter-group interaction.
Note that in the non-interacting scenario, we address the two NV groups with mismatched microwave driving strengths, in order to reduce the Ising term to zero and decouple the two groups.
Intuitively, this additional robustness may reflect the fact that the two groups oscillate collectively due to the inter-group interaction, which provide a larger energy barrier that prevents the uncorrelated flips of the spins in each individual NV group.

\emph{Conclusion and Outlook}--- Our results have presented the first experimental evidence that long-lived robust DTQC can be realized in quasi-periodically driven many-body quantum systems. %
Such DTQC order is fundamentally distinct with those achieved using conventional Floquet Hamiltonian, greatly expanding the landscape of possible non-equilibrium phases of matter.

Looking forward, our work opens up a few intriguing directions.
For instance, the existence of long-lived DTQC in our 3D dipolar interacting NV ensemble relies on slow critical thermalization.
However in this frame work, the lifetime of DTQC cannot be extended to arbitrarily late times.
To overcome this, one can construct many-body localized or prethermal DTQC using synthetic quantum platforms such as cold atoms and superconducting qubits \cite{kyprianidis2021observation,zhang2017observation,zaletel2023colloquium} as well as spin defects in lower dimensions \cite{davis2023probing,PhysRevX.13.041016,gong2023coherent, randall2021many,gong2024isotope}.
Moreover, it would be interesting to explore, both theoretically and experimentally, the existence of other novel non-equilibrium phases of matter beyond DTQC in driven many-body quantum systems.

\bibliographystyle{naturemag}
\bibliography{ref.bib}

\end{document}


\title{Supplementary Material:\\Experimental Realization of Discrete Time Quasi-Crystals}

\author{Guanghui~He,$^{1,*}$
Bingtian~Ye,$^{2,*,\dag}$
Ruotian~Gong,$^{1}$
Changyu~Yao,$^{1}$
Zhongyuan~Liu,$^{1}$\\
Kater~W.~Murch,$^{1,3}$
Norman~Y.~Yao,$^{2}$
Chong~Zu,$^{1,3,4,\dag}$
\\
\normalsize{$^{1}$Department of Physics, Washington University, St. Louis, MO 63130, USA}\\
\normalsize{$^{2}$Department of Physics, Harvard University, Cambridge, MA 02138, USA}\\
\normalsize{$^{3}$Center for Quantum Leaps, Washington University, St. Louis, MO 63130, USA}\\
\normalsize{$^{4}$Institute of Materials Science and Engineering, Washington University, St. Louis, MO 63130, USA}\\
\normalsize{$^*$These authors contributed equally to this work.}\\
\normalsize{$^\dag$To whom correspondence should be addressed: bingtian\_ye@g.harvard.edu, zu@wustl.edu}\\
}

\date{\today}
\maketitle
\tableofcontents
\section{Experimental setup}

We characterize the dynamics of NV centers using a home-built confocal laser microscope. A $532~$nm laser (Millennia eV High Power CW DPSS Laser) is used for both NV spin initialization and detection. The laser is shuttered by an acousto-optic modulator (AOM, G$\&$H AOMO 3110-120) in a double-pass configuration to achieve $>10^5:1$ on/off ratio. An objective lens (Nikon Plan Fluor 100x) focuses the laser beam to a diffraction limited spot with diameter $\sim 0.5~\mu$m, and collects the NV fluorescence. The fluorescence is then separated from the laser beam by a dichroic mirror, and filtered through a long-pass filter before being detected by an avalanche photodiode (Thorlabs). The signal is processed by a data acquisition device (National Instruments USB-6343). The objective lens is mounted on a piezo objective scanner (Physik Instrumente PD72Z1x PIFOC), which controls the position of the objective and scans the laser beam vertically. The lateral scanning is performed by an X-Y galvanometer (Thorlabs GVS212).

The NV centers are created randomly in the sample, so there exist 4 different crystalline orientations of the spin defects. To isolate one group of NV centers, we position a permanent magnet near the diamond to create an external magnetic field $\mathrm{B}\sim 324~$G along one of the NV axes. 
Under this magnetic field, the $|m_s = \pm1\rangle$ sublevels of the aligned NV group are separated due to Zeeman effect, and exhibits a splitting $2\gamma_e B$, where $\gamma_e = (2\pi)\times2.8~$MHz/G is the gyromagnetic ratio of the NV electronic spin.
A resonant microwave drive with frequency $(2\pi)\times 1.963~$GHz is applied to address the NV transition between $|m_s=0\rangle \Longleftrightarrow |m_s=-1\rangle$ sublevels and isolate an effective two-level system.
We note that at this magnetic field, after few microsecond of laser pumping, the associated spin-1 $^{14}$N nuclear spin of the NV center is highly polarized to $|m_I = +1\rangle$ via the excited state level anti-crossing (esLAC) \cite{jacques2009dynamic, fischer2013optical, zu2014experimental}.

The microwave driving field is generated by mixing the output from microwave sources (Stanford Research SG384 and SG386) and arbitrary wave generators (AWG, Chase Scientific Wavepond DAx22000). 
Specifically, a high-frequency signal at $(2\pi)\times 1.838~$GHz from the microwave source is combined with a $(2\pi)\times 0.125~$GHz signal from the AWG using a built-in in-phase/quadrature (IQ) modulator, so that the sum frequency at $(2\pi)\times 1.963~$GHz is resonant with the NV $|m_s=0\rangle \Longleftrightarrow |m_s=-1\rangle$ transition. 
By modulating the amplitude, duration and phase of the AWG output, we can control the strength, rotation angle and axis of the microwave pulses.
To realize DTQC, we perform spin rotation at quasi-periodic times $n_1\tau_1$ and $n_2\tau_2$, where $\varphi = \tau_2/\tau_1 = (\sqrt{5}+1)/2\approx1.61803...$. Note that the interaction time $\tau_1$ and $\tau_2$ are on the order of microseconds, while the temporal resolution of the AWG has a limit of $0.5~$ns. As a result, we specify $\varphi$ to be accurate to $3$ decimal places, $1.618$.
The microwave signal is then amplified by a high-power amplifier (Mini-Circuits ZHL-15W-422-S+) and delivered to the diamond sample through a coplanar wave guide. The microwave is shuttered by a switch (Minicircuits ZASWA-2-50DRA+) to prevent any potential leakage. 
All equipment are gated through a programmable multi-channel pulse generator (SpinCore PulseBlasterESR-PRO 500) with $2$~ns temporal resolution.

\section{Sampling scheme for probing spin dynamics}
Due to the presence of quasi-periodic rotations in the microwave pulse sequence, there is no stroboscopic time to probe the spin polarization. There are two alternative sampling schemes. One is to measure the spin dynamics at a much faster frequency than the rotation frequency, such that one can capture the micromotion between rotation pulses. In practice, we choose the sampling frequency to be $\omega_s = 10\omega_1$, and the probed spin dynamics is shown in Fig.~\ref{fig:DenseSampling}a in yellow. A second possible approach is to probe the average polarization after each rotation pulse (blue in Fig.~\ref{fig:DenseSampling}a), or sampling at the quasi-periodic time. As a consequence, the acquired data points are not evenly spaced in time space, and thus one needs to perform nonuniform Fourier transform to extract the subharmonic response frequencies.

Here, we verify that by sampling at the quasi-periodic time, $t=n_1\tau_1$ and $t=n_2\tau_2$, the measured spin dynamics and the corresponding Fourier spectrum agrees well with their dense sampling counterparts. Shown in Fig.~\ref{fig:DenseSampling}a, the measured spin dynamics by sampling at discrete times overlaps with the dense sampling one. The corresponding Fourier spectra also faithfully provides the subharmonic response frequencies. In Fig.~1-3 in the main text, we probe the spin dynamics with the quasi-periodic sampling scheme.
\figDenseSampling

However, for $\mathbb{Z}_2\times \mathbb{Z}_2$ DTQC, we find that when probing the spin dynamics with quasi-Floquet frequencies, the extracted Fourier spectrum exhibits multiple spurious peaks. Fig.~\ref{fig:DiscreteSampling} shows the measured spin dynamics in time and frequency space with $\tau_1 = 1.00~\mu$s and $1-\epsilon=1$. Except from the expected four subharmonic response frequencies, there exist additional peaks at $\{N\omega_2-\frac{\omega_1}{2}, N\in \mathbb{Z}\}$ related to the discrete sampling scheme (yellow dashed line). To accurately reflect the essence of $\mathbb{Z}_2\times \mathbb{Z}_2$ DTQC, we utilize the dense sampling scheme in Fig.~4b-c in the main text. 
Nevertheless, when we compare the crystalline fraction at subharmonic frequency $\frac{1}{2}\omega_1$ in Fig.~4d in the main text, we adopt the quasi-periodic sampling scheme since the spurious peaks do not affect the position of the subharmonic peaks.
\figDiscreteSampling

\section{Numerical simulation}
The numerical simulation is implemented using Krylov subspace eigensolving via Dynamite~\cite{greg2023dynamite}. Specifically, we track the polarization dynamics of one NV spins surrounded by 15 randomly positioned spins. We only use the polarization from the center spin as a representation of the total spin dynamics, since far away spins suffer from finite-size effect. The dipolar interaction Hamiltonian is built up using Eqn.~2 in the main text. The bath spin density is set to be $\rho_{\frac{1}{4}}=1.125~$ppm. The subscript denotes $\frac{1}{4}$ of the total NV density in our sample, since we only perform experiments on one of the four NV groups. We generate a Gaussian distributed onsite random field, $h_zS^z_i$, with a standard deviation of $1.7~$MHz, which agrees with the experiments. For each simulation, we average 50 realizations of positional disorder and random field.

\section{Optically detected magnetic resonance spectrum}
To create DTQC with $\mathbb{Z}_2\times \mathbb{Z}_2$ symmetry, we need to address two NV groups individually with distinct microwave frequencies. We misalign the external magnetic field, such that the magnetic line is not in parallel with any NV axis. In this way, we isolate two NV groups while separating them from other groups. Also, we carefully choose the B field orientation, to minimize the consequence of the transverse magnetic coupling for both groups.

Fig.~\ref{fig:ESR} shows six peaks in the optically detected magnetic resonance (ODMR) spectrum, indicating the transition frequencies of NV centers. In experiments, we choose NV 1 $|m_s = 0\rangle \Longleftrightarrow |m_s = -1\rangle$ transition as group A, and NV 2 $|m_s = 0\rangle \Longleftrightarrow |m_s = +1\rangle$ as group B.
\figESR

\section{Pulse sequence for the $\mathbb{Z}_2\times \mathbb{Z}_2$ DTQC}
In Fig.~4a in the main text, we display the experimental pulse sequence for observation of the $\mathbb{Z}_2\times \mathbb{Z}_2$ DTQC. However, in real experiments, we need to take finite rotation pulse duration into consideration. Specifically, at time $n_2\tau_2$, we apply two additional pulses on group B, [$\frac{(1-\epsilon)\pi}{2}_{\hat{x}}$~---~$\frac{(1-\epsilon)\pi}{2}_{-\hat{x}}$]. The two adjacent pulses are along the exact opposite axis, so that the effects of the two cancel. Meanwhile, the total time duration of the two pulses matches that of the applied rotation pulse on group A, which compensates the time lapse that would otherwise be a redundant $(1-\epsilon)\pi_{\hat{x}}$ pulse for group B (Fig.~\ref{fig:PulseSeq}).
\figPulseSeq

\section{Inter-group interaction between two NV groups}
In the main text, we utilize two crystallographic groups of NV to realize a more complex DTQC with $\mathbb{Z}_2\times \mathbb{Z}_2$ symmetry, and investigate the effect of inter-group interaction on the robustness of the phase boundary. Here, we verify the existence of the interaction between two groups.

Fig.~\ref{fig:Int} displays the spin locking contrast reduction, induced by the inter-group interaction. First, we apply a spin locking sequence on group B, with a fixed duration $\tau = 0.3~$ms and microwave strength $\Omega=(2\pi)\times8.33~$MHz. Then, we apply a static microwave drive on group A with the same strength $\Omega$, while sweeping the microwave frequency. The remaining polarization of group B is measured. Fig.~\ref{fig:Int}a shows the spin locking amplitude as a function of the microwave frequency. A clear resonance is observed when the microwave frequency matches $|m_s = 0\rangle \Longleftrightarrow |m_s = -1\rangle$ transition frequency of group A, indicating a solid coupling between two groups. We further corroborate this result by performing spin locking on group A while sweeping static microwave frequency on group B. Fig.~\ref{fig:Int}b indicates that the polarization shows a dip when the microwave frequency is in resonance with the $|m_s = 0\rangle \Longleftrightarrow |m_s = +1\rangle$ transition frequency of group B.
\figInt

\newpage
\bibliographystyle{naturemag}
\bibliography{ref.bib}